# SafeZone: A Hierarchical Inter-Domain Authenticated Source Address Validation Solution

Jie Li[a,*], Jian-ping Wu[a], Ke Xu[a]

[a] *Department of Computer Science and Technology, Tsinghua University, Beijing 100084, China*

**Abstract**

Next generation Internet is highly concerned about the issue of reliability. Principally, the foundation of reliability is authentication of the source IP address. With the signature-and-verification based defense mechanisms available today, unfortunately, there is a lack of hierarchical architecture, which makes the structure of the trust alliance excessively flat and single. Moreover, with the increasing scale of the trust alliance, costs of validation grow so quickly that they do not adapt to incremental deployment. Via comparison with traditional solutions, this article proposes a hierarchical, inter-domain authenticated source address validation solution named SafeZone. SafeZone employs two intelligent designs, lightweight tag replacement and a hierarchical partitioning scheme, each of which helps to ensure that SafeZone can construct trustworthy and hierarchical trust alliances without the negative influences and complex operations on de facto networks. Extensive experiments also indicate that SafeZone can effectively obtain the design goals of a hierarchical architecture, along with lightweight, loose coupling and "multi-fence support" and as well as an incremental deployment scheme.

## 1. Introduction

An innate deficiency of the Internet is its susceptibility to IP spoofing. When the source addresses of IP packets being forwarded through a router are not validated, it makes it easy for attackers to spoof a source IP address other than the accurate address of the attacked host. As a result, source IP address spoofing has plagued the Internet for many years. Attackers forge source addresses to trigger attack events and redirect blame. A research report by US CERT [1] shows that the rate of Internet security attacks is increasing much faster than the development of the Internet itself. Attacks such as DoS/DDoS, Botnet and SPAM are frequent occurrences on the Internet, paralyzing Internet services and causing millions of dollars of financial loss. The largest observed attack reached 49Gbps in 2009, a 104% growth over the past two years [2]. Because tens of gigabits of flooding traffic could easily overwhelm most links, routers, or sites on the Internet, DoS attacks have already ranked as one of the largest anticipated threats. Today, challenge of validating the authenticity of source IP addresses is challenging the secure operation and sustainable development of the Internet. With varying levels of success, researchers have proposed many source address validation mechanisms to defend against spoofing, among which classical examples include SPM [3] and APPA [4]. SPM presents an anti-spoofing method which is effective and also supports incremental deployment. However, SPM is not DoS resilient in the key-update process and cannot be used for anti-spoofing of smaller granularities. APPA derives from and improves upon SPM. It uses the same tagging and key verifying mechanism as SPM and makes the key-update process automatic; APPA is more effective and secure. But, when applied to larger scales, the flat and single architecture of the trust alliance (TA) leads to increasing overhead on processing and storing of the state machine (SM), which makes APPA difficult to adapt to an incremental deployment scheme. Based on the two previous methods, this paper introduces a new anti-spoofing solution: a hierarchical, inter-domain authenticated source address validation solution, called SafeZone (§3).

---

\* Corresponding author. Tel.: +086-10-62603064.
*E-mail address*: jieli@csnet1.cs.tsinghua.edu.cn.

SafeZone can work more efficiently in de facto networks without any changes to the topological structure and inter-domain routing protocol(s) under the precondition of ensuring inter-domain high-speed communications. Moreover, SafeZone hardly suffers from changes in the network topology and the route without special operation at the intermediate nodes. Unlike SPM and APPA, all autonomous systems (AS) that deploy SafeZone are divided into a certain number of multilevel TAs; each level of which can serve as a member (abstracted as a whole system) of a higher level TA from bottom to top, the collection of which forms a hierarchical architecture and allows several multi-level TAs to coexist.

By means of the trust alliance border (TAB), which takes the role of a lightweight tag replacement relay agent, SafeZone makes the internal network environment trustworthy and invisible to one another among TAs lying at different levels. This has the effect of eliminating the mutual influence coming from TAs at different levels. Via comparison with SPM and APPA, it is obvious that SafeZone further efficiently reduces the overhead on computing, storing, searching, synchronizing and processing the SM (§4.1). Experimental results show that, even in a larger scale hierarchical TA architecture, SafeZone can still guarantee effectiveness and simplicity (§4.2). SafeZone also establishes and maintains a "deploy early and benefit rapidly" incentive mechanism.

The rest of this paper is organized as follows: Section 2 discusses related work; Section 3 describes the principle underlying SafeZone in detail; Section 4 presents the experimental data and evaluation of those results on the costs of running SafeZone and demonstrations of its efficacy; finally Section 5 concludes the paper.

## 2. Related Work

For many years the research community has been committed to combatting IP spoofing. Existing approaches to handle forged IP source addresses, such as [4],[5],[6],[7],[8], can be broken down into three categories: traceback, filtering en route, and using encryption authentication. Traceback usually requires expensive and complicated techniques to observe traffic as these traffic passes through routers and then to subsequently reconstruct the entire path taken by the packet from source to destination. Moreover, traceback is typically performed after an attack is detected, which does little to console the victim whom has already been damaged.

Filtering en route, the second category of anti-spoofing methods, is a kind of method that actively tries to forged filter packets before arriving at their destination. Typical examples include [9], [10], [11] and [12]. Ingress Filtering brings such benefits to the network only if deployed on a large scale, which makes it an unattractive option for incremental deployment. uRPF is relies on knowing ahead of time a predestined route on which packets destined for a given AS would travel. uRPF also suffers from asymmetry when routing. DPF supports incremental deployment, but it requires an expansion of the BGP, and the variation of the route undertaken by packets may lead to problems of false positives. The SAVE protocol has the router (or set of routers) in charge of a given source address space send SAVE updates corresponding to each forwarding table entry. However, the variation of the route may have validation rules update frequently, which may bring an increasing communication overhead.

Using encryption authentication is another clever way of defending against spoofing packets, such as IPsec, SPM and APPA. IPsec is effective at preventing an attacker from interfering with an existing connection as well as preventing an attacker from successfully initiating a connection. However, IPsec is ineffective at mitigating a bandwidth-based DoS attack, and may in fact exacerbate the problem. In SPM, routers at the edge of the source networks mark keys on outgoing packets; these keys are then verified when they are received at the routers at the edge of the destination. Although SPM is useful in many situations to prevent spoofing, the interaction of key updating could easily become the target of DoS/DDoS attacks. Moreover, the costs of encrypting, consulting, and synchronizing between AS pairs preclude the ASes from maintaining many active key-related communication connections. Whereas the original SPM could not be deployed on a large scale, APPA makes each key valid only in one packet and subsequently produces another, entirely new key for each of the following packets. Key updating is finished automatically by special mechanism rather than interaction. This improves APPAs safety and efficiency making eavesdropping and reuse of the key or the whole packet useless. However, due to all member ASes belong to one single TA, the structure of TA is constructed to be merely flat and single. Moreover, considering the two-way orderliness of the SM, in order to ensure the efficiency of the validation mechanism, all member ASes must establish a full-mesh and two-way shared relationship among each of the AS pairs corresponding to the SM, which results in all border routers residing on every member AS having to maintain the total (large) number of global SMs. In the extreme case, when the scale of member ASes amounts to $N$, each border router must maintain as many as $2(N-1)$ SMs causing the space complexity to reach $O(N)$. The cost of SM search and storage, consultation and synchronization are heavy and difficult for routers to validate correctly and control at an efficient level. With the increase in scale of TA, any change coming from the member AS can influence the whole TA situation. Therefore, these aforementioned problems may be a barrier for incremental deployment. Unfortunately, when the scale of TA increases to a certain scale, APPA may even be inefficient and unreliable.

## 3. Design principle of the SafeZone

In this section we describe a new inter-domain authenticated source address validation solution, SafeZone. SafeZone advocates an intelligent, light-weighted tag replacement based mechanism without negative influence on real-world networks and can provide a way to build a hierarchical architecture. This architecture presents a novel multi-level TA structure. We first summarize a list of technical terms used throughout this paper in Table 1.

Table 1. A list of technical terms

| Technical Terms | Abbr. | Notes |
|---|---|---|
| Trust Alliance | TA | The set of members deploying validation rules |
| Registration Server | RES | The server used for member registration and management ( 1/TA ) |
| AS Control Server | ACS | The server used for communicating with RES and ACSes residing on other member ASes and configuring all ABRs/TABRs residing on local AS ( 1/member AS ) |
| AS Border Router | ABR | The AS border router with the ability of source address validation ( > 1/ member AS ) |
| TA Border | TAB | The member residing on local TA and connecting local TA to outside network ( designated beforehand ) |
| TA Border Router | TABR | The border router residing on local TAB and connecting local TA to outside network, which has the ability of source address validation ( > 1/ TAB ) |
| State Machine | SM | The method used for generating the Tag at source and verifying the Tag at the destination |
| Tag | Tag | The string generated by SM and added into each packet at the source and verified at the destination |
| Global TA SM | GSM | The SM used globally for members belonging to different level TAs |
| Member and TA Mapping SM | MSM | The SM used in cross-TA communication (§ 3.3), which denotes a mapping relationship between one TA and its member |

### 3.1. Basic Rationale

As a light-weighted signature-and-verification solution, SafeZone makes all ASes who deploy SafeZone construct a hierarchical organization and allows several multi-level TAs to coexist. That is, all ASes are first divided into a certain number of "lowest level" TAs, allowing each "lowest level" TA to join other higher level TAs as a member (abstracted as a whole system). Continuing in this fashion, the lower level TAs can continuously combine into several higher level TAs from bottom to top, finally forming a hierarchical architecture ( § 3.2). Through the work of the TAB, which takes charge of the lightweight tag replacement as the bridge or relay agent, SafeZone makes the internal network environment self-visible to one another among different level TAs, eliminating the mutual negative influence of change. SafeZone focuses on structuring the hierarchical TA architecture and ensuring the functionality of source address validation without negative influence and complex operations on de facto networks.

### 3.2. Architecture

In accordance with real world network situations, SafeZone enables the supervising organization of TA to flexibly construct the hierarchical TA through the ability to choose different strategies such as classification principles and combination models. At first, the ASes with equivalent situations are divided into several lowest level TAs, where a single AS is regarded as the unit member. Then, these lowest level TAs are further aggregated to form several higher level TAs on the basis of certain classification principles, where the single lowest level TA is regarded as the unit member, and so on. In this way, the TAs lying in the same level are continually combined to be contained in a certain (smaller) number of higher level TAs in a bottom-up fashion until they become one ultimate, "highest level" TA. Finally, this presents a hierarchical nested structure as shown in Figure 1. In the architecture of SafeZone, the main equipment used are border routers which takes part in tag adding, replacing and verifying. Considering their respective functions, there are two general types of border routers: ABR and TABR. The ABR resides on the lowest level TA, which is in charge of firstly adding and then finally verifying the tag, and does not consider tag replacement or the global situation of TA. The ABR just maintains the SM in the scope of the lowest level TA. Note that the ABR's number is equal to the amount of member ASes in the lowest level TA. The TABR allocated on the TAB is the crucial link between TAs lying at different levels. The TABR is responsible for the implementation of tag replacement in the cross-TA communication scenario ( § 3.3). Because the process of source address validation involves several multiple level TAs and requires collaboration at every level TA, the TABR inside each level TA needs to maintain the GSM.

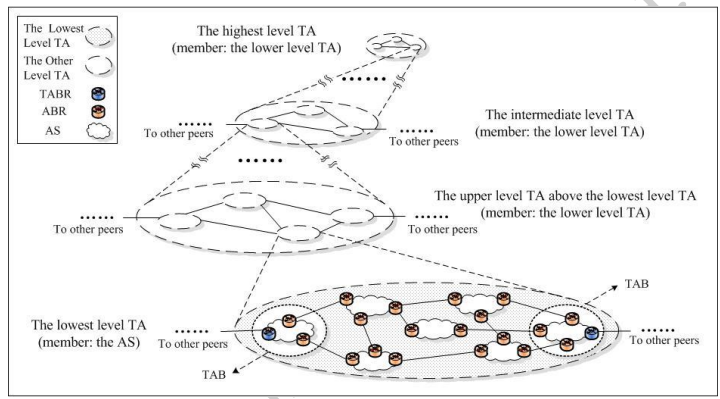

Fig. 1. The hierarchical architecture of SafeZone

*3.3. Data Plane*

In the hierarchical architecture of TA, through level division, data communication is expanded into three categories, each with different communication scenarios. These three communication scenarios of the data platform can be outlined as follows:

1. Single TA Communication. This first scenario denotes member ASes belonging to one of the lowest-level TAs mutually communicate among one another. In addition, packets are delivered in the scope of the local network of the lowest-level TA. In this scenario, referring to SMs covering the lowest-level TA, the ABRs only need to execute the traditional mechanism to effectively achieve source address validation. The process of validating packets obviously involves no tag replacement.

2. Cross-TA Communication. In this second scenario, the ASes belonging to different TAs mutually communicate among one another without establishing a full-mesh and a two-way shared relationship is formed corresponding to the SM by means of introducing the TAB and the MSM. At first, packets sent from the source AS are marked with a tag by the source ABR according to the corresponding MSM, guaranteeing the initial authenticity. When packets are transfered from the source to the first TAB en route, the TABR can implement tag replacement with the relevant MSM and GSM. The tags marked in packets are replaced with new ones corresponding to the higher level TA along the actual route. As long as the packets transit several higher level TAs, the TABRs allocated to these TAs continue to complete the tag replacement in this way from bottom to top. During this process, therefore, the serials of TABRs serve as a connecting bridge, or relay agent, in the communication scenario, ensuring the authenticity of the source address among different level TAs. With packets reaching TAs close to the destination AS, on the other hand, the TABR belonging to each level TA can carry out reverse operations from top to bottom. When packets are forwarded to the TAB of the lowest level TA close to the destination AS, the TABR finally replaces the tag referring to the MSM. During the process of forwarding packets en route, other intermediate nodes do not need to deal with the tagged packets. Finally, when the packets reach the destination AS, the ABR verifies and removes the tag added to the packets by checking the MSM, accomplishing the whole process of source address validation. This process is illustrated in Figure 2.

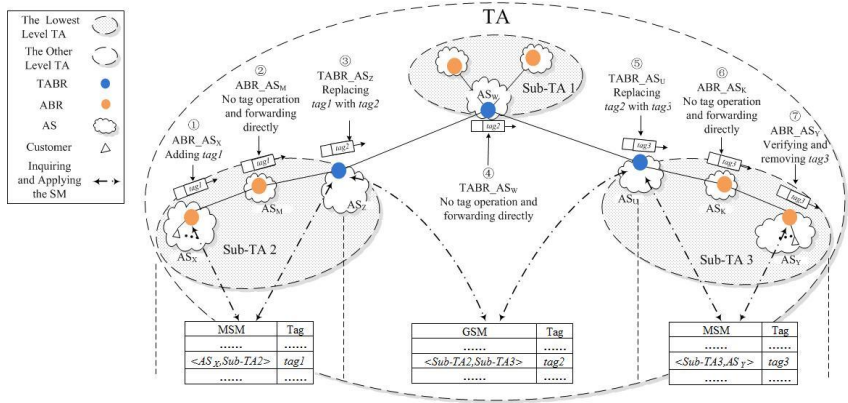

Fig. 2. The process of tag replacement in SafeZone (Cross-TA data Communication Scenario)

3. TA-Member AS and Non-TA-Member AS Communication. In this scenario, the communication is between the TA-member AS and the non-TA-member AS who does not deploy SafeZone. In addition, packets are just forwarded along the path without any operations on their tags. This scenario does not involve any source address validation.

*3.4. Control Plane*

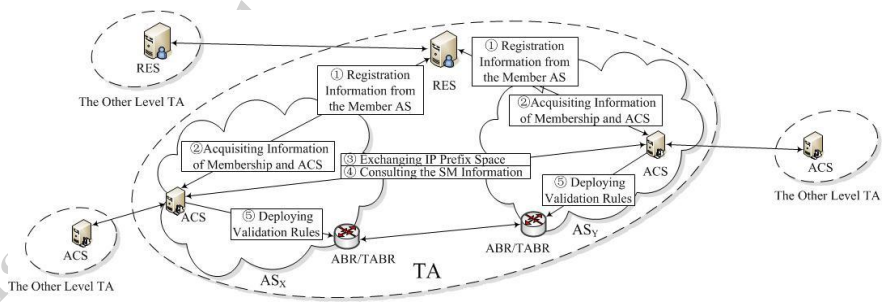

Fig. 3. Illustration of the Control Plane

In control plane, SafeZone requires that the ABR/TABR makes a concerted effort with RES and ACS to accomplish tasks including consultation, decision and deployment on validation rules. In the hierarchical architecture, each level TA needs to deploy one RES to be responsible for two tasks. The first task is to control in a downward fashion, the joining and the leaving of members, the advertizing of local membership, and the ACSes' address space information to local ACSes allocated in the same level TA. The second task is to exchange in an upward fashion, information about membership and ACSes with other RESes residing on other, higher level TAs.

While RES is deployed, each member AS needs to deploy one ACS. The ACS is responsible for three tasks. The first is to register with the local RES and to acquire information about membership as well as other ACSes' address space. The second task of the ACS is to exchange the IP prefix space and to consult the initial SM with other ACSes. The third and most important task is to set up the mapping table between the IP prefix and AS in addition to establishing the two-way SM table on ABR/TABR and then subsequently taking control of them to successfully execute the duties on the data plane. Simultaneously, the ACS also communicates upward with other ACSes allocated to other, higher level TAs to complete similar work. It is important to remember that during the implementation process, every TAB makes known their status of being a relay agent as well as the ACS residing on the TAB can finish the configuration on TABR. The work process of the control plane is shown in Figure 3.

*3.5. Security Concerns*

For the sake of preventing eavesdropping and the interception of validation information, we specially take account of the robust design as follows: (1) The SM periodically auto-updating the scheme. The scheme focuses on introducing activation time and expiration time for each SM. When the SM expires, both the source and destination AS must auto-update the SM simultaneously so that it can guarantee three characteristics, namely: timeliness, uniqueness and reliability. This is fast and easy. (2) The time synchronization scheme. In SafeZone, in order to solve the problem of clock-drift, RES can use the Network Time Protocol (NTP) for synchronizing the clocks of ACSes by regularly delivering time correction packets allowing each ACS to correct the clock of relevant ABR/TABR. In addition to this, ACS triggers a shared time slice during which both the exactly expired tag and the exactly updated tag are treated as valid tags for source address validation. Consequently, we can further ensure synchronization of the SM. (3) The secure channel scheme. Under de facto inter-domain networks, we employ a combination of TCP interception and the Diffie-Hellman protocol in order to set up a secure channel for the safe exchanging of validation information through the established TCP connection. (4) The working modes of TABR. According with ascertaining the security situation of network in a timely manner, in our designs we allow TABR to switch operating modes flexibly during the process of tag replacement. The operating modes are divided into the mode of direct replacement and the other is the mode of replacement as well as validation.

*3.6. Properties of SafeZone*

Compared with the flat, inter-domain source address validation methods proposed in [3, 4], we are in a position to further define the desirable properties of the SafeZone as follows:

- Hierarchical architecture: By the means of partitioning, SafeZone creates a multi-level coexistence of TAs and constructs a hierarchical architecture as opposed to that of the flat, single structure.
- Internal independence among different level TAs: Depending on the work of the TAB, which serves as lightweight tag replacement relay agent, SafeZone can make the internal network environment trustworthy and invisible among different level TAs as well as preventing influences between each other.
- Low overhead: SafeZone can considerably reduce overhead on computing, storing, searching, synchronizing and processing the SM in cross-TA communication scenario without establishing a full-mesh and two-way shared relationship corresponding to the SM.

Experimental results show that, even in large scale hierarchical TA architectures, SafeZone can still guarantee effectiveness and simplicity. The degree of optimization on the costs of validation is observable (§4.2).

**4. Analysis and Evaluation**

As we already indicated in Section 3, the number of the SMs directly affects the overhead on computing, storing, searching, synchronizing and processing. The number of the SMs is a critical factor to be evaluated. In this section, we analyze the effectiveness of SafeZone in reducing the number of SMs. We then evaluate the optimization of the number of SMs based on an experimental deployment.

## 4.1. Efficiency Analysis

In this section, we prove that the hierarchical structure is effective in reducing the number of the SMs maintained in ABR/TABR. In a flat structure, we assume that *N* denotes the number of ASes, while the number of the SMs each ABR maintained is $S_{(f)ABR}=2(N-1)$, corresponding to the set $C_F$. For a TA *TA(N,L)* in SafeZone, *N* corresponds to its member ASes and *L* to its level ($1<L<\lfloor log_2 N \rfloor+1, L \in Z$). We let $m_i$ represent the total number of members in the *i*-th level TA ($1<m_i<\lceil N/3 \rceil, m_i \in Z$ and uniform), and $p_{ij}$ represents the total number of members in a sub-TA *Sub-TA(N,L)$_{ij}$* (here the *j*-th sub-TA in the *i*-th level in *TA(N,L)* ($1 \leq i \leq L, 1 \leq j \leq m_i, 1<p_{ij}<\lceil N/3 \rceil, i,j,p_{ij} \in Z$)). Note that for demonstrating the hierarchical structure, we suppose that *TA(N,L)* is abstracted to a *L*-height tree $\mathbb{T}$, where each node represents one certain level TA and the number of all leaf nodes is the number of ASes who have deployed SafeZone (Figure 4). We also define $S_{(h)ABR}$ to be the sum of the SMs each ABR maintained corresponding to the set $C_{ABR}$ and $S_{(h)TABR}$. The sum of the SMs each TABR maintained corresponds to the set $C_{TABR}$.

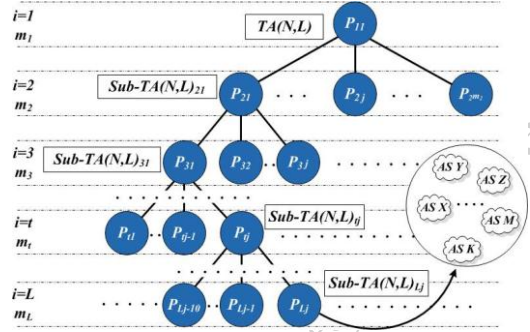

Fig. 4. The hierarchical structure of tree $\mathbb{T}$ (It reflects the hierarchical relation among multi-level TAs and the component of each level TA)

1. In the uniform hierarchical structure constructed by SafeZone, we can effectively reduce the costs of storing and maintaining the SM.

In our definition, "uniform" means: every level TA belonging to *TA(N,L)* has the same scale of *m* members. Here, $\mathbb{T}$ becomes a full *m*-tree, whose height is *L* and includes *N* leaf nodes. First, we deduce the following equation: $S_{(h)ABR}=2(\lceil N/m^{L-1} \rceil-1)+2$. Considering $L<<N$ and $m<<N$ ($N,m \in Z$), we can see that $\lceil N/m^{L-1} \rceil << N$. Then, we can further deduce $S_{(h)ABR}=2(\lceil N/m^{L-1} \rceil-1)+2 << S_{(f)ABR}=2(N-1)$. Second, the number of the SMs maintained by TABR consists of two parts: one, the SMs corresponding to local AS with its peers and the other, those SMs corresponding to local TA with its peers. In general, we can conclude that if several different level TAs overlap in TAB, the TABRs TABR$_{max}$ maintains the largest number of SMs corresponding to the set $S_{(h)TABR \max}$ as shown in equation: $S_{(h)TABR \max}=2\lceil N/m^{L-1} \rceil+2(L-1)m-4L$. According to the hierarchical structure and tag replacement mechanism described in Section 3, the value of *L* should be constrained in favour of not only partitioning and reducing the number of SMs, but also in such a way that avoids excess costs on tag replacement. The value of *m* should also be no surge and uniform. Finally, it is not difficult to deduce that there is always at least one group $\{L, m\}$ that minimizes $S_{(h)TABR \max}$ with increasing of *N*. Then, by taking the derivative of the equation $S_{(h)TABR \max}=2\lceil N/m^{L-1} \rceil+2(L-1)m-4L$, we get that when $m=N^{1/L}$, $S_{(h)TABR \max}$ achieves a minimal value. Since $1<L<\lfloor log_2 N \rfloor+1$, we arrive at the conclusion $S_{(h)TABR \max}=2L(N^{1/L})-4L << 2(N-1)$. Based on the principle of set theory, these results also prove that for any $ABR \in TA(N,L)$ or $TABR \in TA(N,L)$, $C_{ABR} \subset C_F, C_{TABR} \subset C_F$ is rational.

2. In the non-uniform hierarchical structure constructed by SafeZone, we can still effectively reduce the costs of storing and maintaining the SM.

In our definition, "non-uniform" means: those TAs existing on the same of different levels belonging to *TA(N,L)* are not exactly the same scale of *m* members. Every TA can flexibly plan its scale by choosing different classification principles and a different combination of models. Under such conditions, it is not difficult to find the following equation: $S_{(h)ABR}=2(p_{Lj}-1)+2$. Firstly, since all $p_{Lj}<<N$ it is then clear that $S_{(h)ABR} = 2(p_{Lj} - 1) + 2 \ll S_{(f)ABR} = 2(N-1)$. Secondly, as analyzed previously, we can conclude the following equation:

$$S_{(h)TABR_{\max}} = 2(p_{Lj}-1) + 2(p_{L-1j}-1) + ... + 2(p_{ij}-1) + ... + 2(p_{11}-1) + 2(p_{Lj}-1) + 2(p_{L-1j}-1) + ... + 2(p_{ij}-1) + ... + 2(p_{11}-1)$$

$$\Rightarrow S_{(h)TABR_{\max}} = 4\sum_{i=1}^{L}(p_{ij}-1), (p_{1j} = p_{11}) \quad (1)$$

According to the validation mechanism described in Section 3, the value of *L* should be constrained within certain scope, in favour of not only partitioning and reducing the number of SMs, but also avoiding the excess costs on tag replacement. The value of $m_i$ and $p_{ij}$ should also be no surge and uniform. Finally, it is not difficult to deduce that there is always at least one group $m_i$ and $p_{ij}$ that achieves (2). Meanwhile, based on principles of set theory, we also prove that for any $ABR \in TA(N,L)$ or $TABR \in TA(N,L)$, $C_{ABR} \subset C_F, C_{TABR} \subset C_F$ is rational.

$$S_{(h)TABR_{\max}} = 4\sum_{i=1}^{L}(p_{ij}-1) < 4\sum_{i=1}^{L}p_{ij} \ll 2(N-1) \quad (2)$$

In summary, compared with [3, 4] and other relative methods, SafeZone is more effective and more safe in reducing the number of the SMs maintained in ABR/TABR. This can further achieve the goal of considerably reducing the overhead on validation. This experiment also indicates that SafeZone can obtain the design goals of hierarchical architecture, lightweight, loose coupling, "multi-fence support" and incremental deployment (§4.2).

*4.2. Experimental Evaluation*

Based on deployment experiments, we evaluate the adventages of SafeZone on the optimization of the number of the SMs over [3, 4]. In both of the below experiments, SafeZone is deployed on a network environment where the TA has either a uniform, hierarchical structure or non-uniform, hierarchical structure, respectively. We start our experiments under the precondition that all RESes and ACSes have already completed the process of exchanging validation information and configurations on ABR/TABR. The routing protocol used in both deployment experiments is the Border Gateway Protocol version 4 (BGPv4).

1. Experiment I: To verify the effectiveness in uniform, hierarchical structure built upon SafeZone.

Network Simulator version 2 (NS2) [27] was used to evaluate the performance of SafeZone in Experiment I. For simulation purposes, the maximum number of member ASes is set to a relatively high 40,000 on the simulation environment in accordance with the report in [28]. According to the uniform hierarchical structure, we select respectively 4 groups: {*L*=4, *m*=5}, {*L*=5, *m*=5}, {*L*=5, *m*=4} and {*L*=4, *m*=4}, to evaluate the performance on optimization on ABR and 4 groups: {*L*=2}, {*L*=3}, {*L*=4}, {*L*=5} to evaluate the performance on optimization on TABR. From the results shown in Figure 4 and 5, we can see that the number of SMs is far less than those in flat structures, and the increasing function of the SMs evolves approximately into logarithmic function ($L*\log_2(2(N-1))$). Specifically, optimization is quite obvious and space complexity is reduced to a feasible level of $O(N^{1/L})$, reduced from the $O(N)$ of previous methods. Meanwhile, the results can demonstrate that even on large scales, deployment of SafeZone still performs better than [3, 4] and other relative methods.

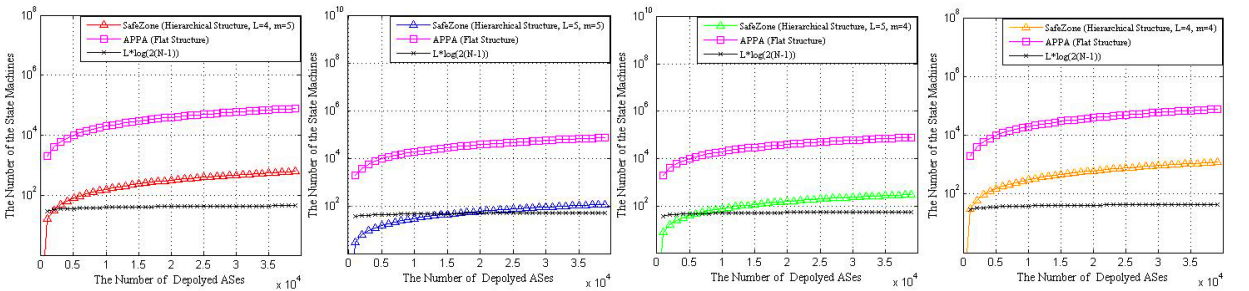

Fig. 5. The number of SMs maintained by ABR in SafeZone

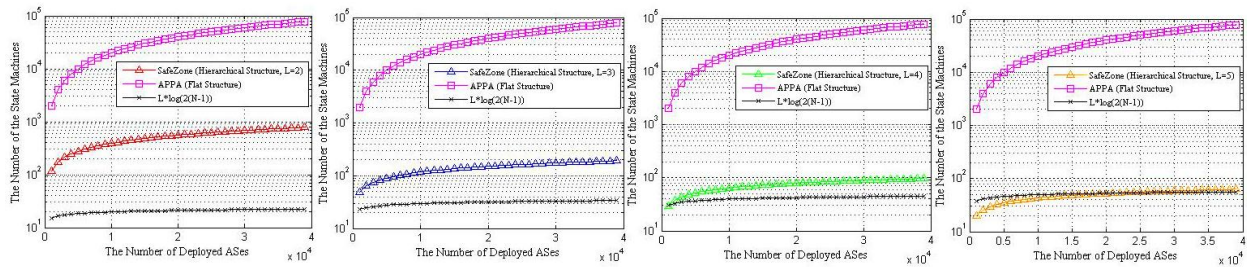

Fig. 6. The number of SMs maintained by TABR in SafeZone

2. Experiment II: To verify the effectiveness in non-uniform, hierarchical structures built upon SafeZone. Supported by the project of the National High Technology Research and Development Program of China (863 Program), Tsinghua University whose aim it was along with other research groups, to address the issue of the weakness on safety and reliability, and to make efforts to solve the difficult problems of authenticated address access. So far, on the basis of the development and implementation of a trustworthy Internet infrastructure, security service and typical application, three IETF standard drafts [13-15] have been submitted; one document eventually became a RFC, RFC5210 [17]. Currently, the prototype system based on authenticated IPv6 source address validation architecture has been deployed and tested on CNGI-CERNET2. With the development of this project, many research institutes, ISPs and equipment manufacturers have become involved, including domestic members CERNET2 [18], ChinaTelecom and ChinaMobile, overseas members including in TEIN3 [19], GÉANT2 [20], APAN-JP [21] and KREONet2 [22]. These members are all designated more than 225 globally unique AS Nums, which provided an ideal network environment for our Experiments II. Based on traffic data collected from Aladdin NMS in NOC of CNGI-CERNET2 we analyzed, we found that the portion of internal traffic within CNGI (blue) by far exceeds the portion of traffic accesses to NGI (red) (In Figure 7 (a)). Also, the portion of internal traffic within CERNET2 (blue) exceeds by a large margin the portion of traffic accesses to CNGI (red) (In Figure 7 (b)). Due to these findings, we constructed the global Next Generation Internet TA shown in Figure 8. This kind of TA resulted in not only the ability to partition and reduce the number of SMs, but also the ability to avoid the excess costs on tag replacement. Compared to [3, 4] and other relative methods, the average rate of reduction can reach to 85% shown in Table II. Our experimental results show that SafeZone is more effective, safe and feasible.

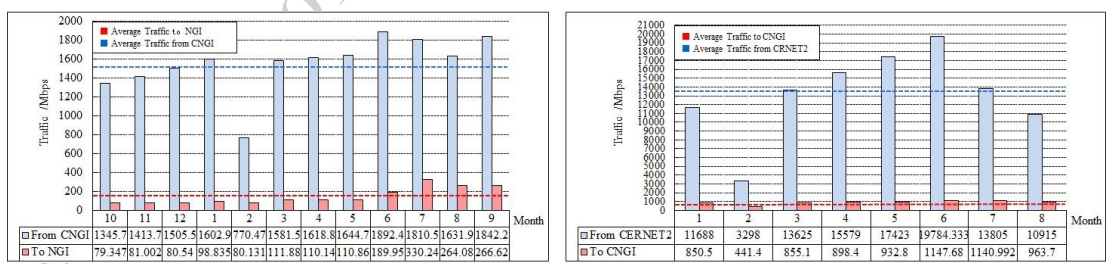

Fig. 7. (a) The traffic statistics in CNGI-6IX (10.2009~9.2010); (b) The traffic statistics in CNGI-CERNET2 (1.2010~8.2010)

| Member | AS | Flat Structure SM (ABR) | Hierarchical Structure | | | |
|---|---|---|---|---|---|---|
| | | | SM (ABR) | Reduction (%) | SM (TABR) | Reduction (%) |
| CERNET2 | 25 | 448 | 48 | 90% | 58 | 87% |

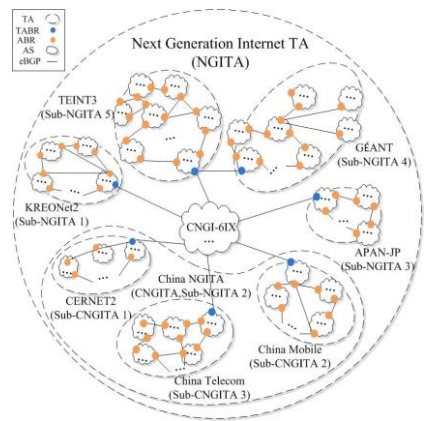

| | | | | | | |
|---|---|---|---|---|---|---|
| ChinaTelecom | 45+ | 448 | 88 | 80% | 98 | 78% |
| ChinaMobile | 55+ | 448 | 108 | 76% | 118 | 74% |
| TEIN3 | 20+ | 448 | 38 | 92% | 48 | 89% |
| GÉANT2 | 34+ | 448 | 66 | 85% | 76 | 83% |
| APAN-JP | 32+ | 448 | 62 | 86% | 72 | 84% |
| KREONet2 | 14+ | 448 | 26 | 94% | 36 | 92% |

Fig. 8. (a) The comparison of the number of SMs between two structures; (b) The hierarchical organization of NGITA

## 5. Conclusion

Compared with [3, 4] and other relative signature-and-verification solutions, SafeZone focuses on ensuring the functionality of source address validation based on de facto networks and has been found in our experiments to especially adept at flexibly constructing the hierarchical architecture of TA. By introducing two intelligent designs: lightweight tag replacement and hierarchical partitioning, SafeZone can make the internal network environment trustworthy and invisible among different level TAs, eliminating the mutual influence coming from these multi-level TAs under the precondition of ensuring inter-domain high-speed communications. Experimental results show that even when implemented on large scale hierarchical architectures, SafeZone can still achieve the effectiveness, simplification, and optimizing of costs on the SM. In future work we will focus on experimental deployment and tests based on CNGI-CERNET2 in preparation of global scale deployment on CNGI. We will conduct further research on designing and improving the performance of SafeZone to make it safer and more light-weight.

## Acknowledgements

The authors would like to appreciate the anonymous reviewers for their comments. This research is supported by the project of the National High Technology Research and Development Program of China (863 Program) under Grant No. 2009AA01A334, No. 2008AA01A323, and No. 2008AA01A326.